\documentclass[fleqn,usenatbib]{mnras}

\usepackage[T1]{fontenc}

\DeclareRobustCommand{\VAN}[3]{#2}
\let\VANthebibliography\thebibliography
\def\thebibliography{\DeclareRobustCommand{\VAN}[3]{##3}\VANthebibliography}

\usepackage{graphicx}	
\usepackage{amsmath}	
\usepackage{amssymb}	
\usepackage{xcolor,colortbl}

\usepackage{newtxtext,newtxmath}

\title[Magnetospheric accretion onto RZ Psc]{Magnetospheric accretion at the late phases of the Pre-Main-Sequence evolution. The case of RZ Psc}

\author[D. V. Dmitriev et al.]{
D.V. Dmitriev, $^{1,2}$\thanks{E-mail: @. (dmitrievdv242@gmail.com)}
T.A. Ermolaeva,$^{1}$
V.P. Grinin$^{1,3}$ and I.S. Potravnov$^{4,5}$ 
\\
$^{1}$Central (Pulkovo) Astronomical Observatory of the Russian Academy of Sciences, Pulkovskoye Chausse 65/1, 196140, St.Petersburg, Russia\\
$^{2}$Crimean Astrophysical Observatory of the Russian Academy of Sciences, p/o Nauchny, 298409, Republic of Crimea\\
$^{3}$St. Petersburg State University, Universitetskii pr. 28, 198504, St. Petersburg, Russia\\
$^{4}$Institute of Solar-Terrestrial Physics, Siberian branch of Russian Academy of Sciences, Lermontov Str. 126A, 664033, Irkutsk, Russia\\
$^{5}$Institute of Astronomy of the Russian Academy of Sciences, Pyatnitskaya str. 48, 119017, Moscow, Russia
}

\date{Accepted XXX. Received YYY; in original form ZZZ}

\pubyear{2015}

\definecolor{RedAlert}{rgb}{1,0.3,0.3}

\begin{document}
\label{firstpage}
\pagerange{\pageref{firstpage}--\pageref{lastpage}}
\maketitle

\begin{abstract}
It has been shown that during theburst of accretion activity observed in UX Ori type star RZ Psc in 2013, the accretion rate increased approximately by an order of magnitude. This means that the accretion process at the late stages of the Pre-Main Sequence evolution is very unstable. Using the spectra obtained during this episode we have studied the magnetospheric emission in the \(\mathrm{H\alpha}\) line. Models of magnetospheric accretion are calculated to obtain the parameters of the magnetosphere from this observation. In present work we have taken into account the influence of the recombination delay effect during gas motion in the stellar magnetosphere. The accounting for this effect and the presence of the magnetospheric absorption in the IR CaII triplet lines and its absence in D Na I resonance lines allowed us to place a lower limit on the temperature in the magnetosphere at $\approx 10000$ K, which significantly improved precision of our estimate of accretion rate. According to the best fit model the logarithm of accretion rate is \(\log\dot{M} =  -10.1\pm0.3\) (\( \dot{M} \approx 7\times10^{-11}\ \mathrm{M_\odot yr^{-1}}\)) and the inclination angle of RZ Psc is $43\pm 3^\circ$. It is less than the inclination, typical for the UX Ori stars (about 70$^\circ$), that explains the weak photometric variability of this star. Using the obtained accretion rate and magnetosphere radius we estimate the strength of the dipole component of the magnetic field of RZ Psc $\approx$ 0.1 kGs.

\end{abstract}




\section{Introduction}

The star RZ Psc (Sp = K0 IV, \cite{herbigSpectralClassifications1121960}) is one of the most unusual members of the UX Ori stars (UXOrs) family. The members of this family are the photometrically most active young objects. They demonstrate the deep (up to 2-3$^m$ in V band) sporadic brightness minima with a typical duration from a few days to a few weeks. The reason for this activity is the complex structure of the nearest circumstellar (CS) environment of young stars. When the direct stellar radiation is blocked by a CS dust cloud crossing the line of sight the scattered radiation of CS disk dominates, and UXOr becomes a highly polarized object (\cite{grininInvestigationsZodiacalLight1991a}). This important observational fact tells us about the small inclination of CS disk planes of UX Ori stars relative to the line of sight as the main reason of their photometric activity.

Unlike typical UXORs the star RZ Psc shows very short Algol-like minima with the typical duration of 1-2 days \citep{zaitsevaMinimaRZPsc1978, pugachPhenomenologicalModelAntiflare1981}. Such short eclipses are similar to those observed in the eclipsing binary systems, although many attempts to find a period were unsuccessful (\cite{kennedyTransitingDustClumps2017} and references therein). 

For a long time evolutionary status of the star was unclear. RZ Psc is not located close to any known star formation regions: its galactic lattitude is high (about 35 degrees) and there are no emission lines in star spectrum \citep{herbigSpectralClassifications1121960, kaminskiiSpectralFeaturesRZ2000}. No infrared (IR) excess was observed in JHK bands. Spectral observations by \cite{herbigSpectralClassifications1121960} did not reveal any signatures of youth.  

The first signs of a dusty disk (or disk-like envelope) have been observed by \cite{kiselevStrongIncreaseLinear1991}. It was found that the linear polarization of the star increased up to approximately 5\% during the deep minimum, that is typical for UXOrs. 
These observations were confirmed by \cite{Shakhovskoi2003PhotometryAndPolarimetryRZPsc}. The latter authors first suggested that RZ Psc is surrounded by circumstellar disk with the central cavity free (or almost free) of matter. This assumption was confirmed by \cite{dewitActiveAsteroidBelt2013}:  a bright mid-IR excess was found in WISE observations of RZ Psc, fitted by black-body radiation with temperature \(\approx500\) K. They assumed that the star is surrounded by the dusty ring with inner radius 0.4-0.7 AU and this assumption was recently confirmed by \cite{kennedyLowmassStellarCompanion2020a}. They discovered that RZ Psc hosts a 0.12 M$\odot$ companion at a projected separation of 23 AU.    

The spectroscopic observations have shown that the Li I 6708 \AA\, line is present in the
spectrum of RZ Psc and has an equivalent width EW(Li) = 0.202 ± 0.010 \AA \, \citep{grininEvolutionaryStatusUX2010}. Using the lithium depletion trend and kinematical treatment of RZ Psc the first age estimate of approximatly 30-40 Myr  was made for the star, that was later refined by \cite{potravnovFlaresAccretionActivity2019} to 20$\substack{+3 \\ -5}$ Myr. This allowed us to reinforce the evolutionary status of RZ Psc as the post UX Ori star \citep{grininEvolutionaryStatusUX2010}. 

So, in terms of accretion activity RZ Psc is a weak lined T Tauri star (WTTS). At the same time, according to de Wit et al. (2013) the star has quite strong mid IR excess ($L_\mathrm{IR} \approx 0.08 L_\mathrm{bol}$) that is not typical for WTTS's.      

The first evidence for existence of a magnetosphere around RZ Psc came from observations of the narrow blue-shifted absorption components (BACs) in the sodium D~NaI lines \citep{grininMagneticPropellerEffect2015}. To explain the origin of these components, the model of magnetospheric accretion in the magnetic propeller mode was used. Estimates have shown that for realisation of this regime, the magnetosphere must be large, extending up to 10 stellar radii. 
The spectroscopic monitoring of RZ Psc revealed occasional presence of the weak emission in the core of the photospheric $\mathrm{H\alpha}$ line \citep{potravnovAccretionOutflowActivity2017}. Interestingly, the $\mathrm{H\alpha}$ emission was detected  when the star was near the bright photometric state. An estimation of the mass accretion rate $\dot{M} \le 7\times 10^{-12}\ \mathrm{M}_\odot\mathrm{yr}^{-1}$ was made using the empirical calibration of  the $\mathrm{H\alpha}$ flux versus accretion luminosity. 

The very interesting spectral observations of RZ Psc  have been made by \cite{punziYoungStarRZ2018} at November 2013, after deep photometric minimum. The $\mathrm{H\alpha}$ line in that spectra demonstrated the classical signs of accretion: the red-shifted absorption component. The accretion rate estimated from this profile using the same empirical method was $\approx 5 \times 10^{-11}\ \mathrm{M}_\odot\mathrm{yr}^{-1}$ \citep{potravnovFlaresAccretionActivity2019}. So, \cite{punziYoungStarRZ2018} observed a burst of accretion activity.

We assume that during the accretion burst a quasistable magnetosphere was formed. That is true, if the duration of the burst \(t_\mathrm{burst}\) was sufficiently larger than the free fall time from the base of the magnetosphere (\(t_\mathrm{ff}\)). The fact, that the redshifted absorption component is observed at large interval of the velocities (from approximately 100 km/s to almost the escape velocity of the RZ Psc \(\approx 600\) km/s) means that the gas have managed to fall onto the star before the burst ended, and thus \(t_\mathrm{burst} \ge t_\mathrm{ff}\). The regular look of the absorption component and the fact, that we were able to reproduce the observed profile supports this assumption.

The goal of our paper is to model the $\mathrm{H\alpha}$ line profile to determine parameters of the magnetosphere: temperature, accretion rate, size and inclination angle of the magnetosphere axis. The model we are using is described in detail in our previous papers \citep{dmitrievFormationHydrogenEmission2019, dmitrievModelsMagnetosphericAccretion2022} and based on the classical approach to modeling of T Tauri stars magnetospheres \citep{hartmannMagnetosphericAccretionModels1994, muzerolleEmissionLineDiagnosticsTauri2001}. However, unlike previously developed models, our model takes into account advective transfer of ionization that can be important in low density plasma when collisional recombination and ionization processes are slow (see section \ref{sec:model}). As it is shown in \cite{dmitrievModelsMagnetosphericAccretion2022}, this effect can be important at the low accretion rates \(\dot{M} \le 10^{-9}\ \mathrm{M}_\odot\mathrm{yr}^{-1}\), like in the case of RZ Psc.

The description of observational data used is given in Section \ref{sec:obs}. Section \ref{sec:model} gives a brief review of our model. The results and details of fitting procedure are given in Section \ref{sec:res+fit} and discussed in Section \ref{sec:disscuss}. A short review of main results is given in Section \ref{sec:concl}. 

\section{Observational data}\label{sec:obs}

In our analysis we used high resolution optical spectrum retrieved from the Keck Observatory Archive\footnote{https://www2.keck.hawaii.edu/koa/public/koa.php}, demonstrative for the accretion activity of RZ Psc. This is essentially the same material which was previously discussed in \citet{punziYoungStarRZ2018,potravnovFlaresAccretionActivity2019}. We recall that accretion is almost ceased in $\sim20$ Myr old RZ Psc system and manifested as the short-term "accretion flares"\,. It is a difficult observational challenge to catch the star during such a sporadic accretion event. Hence, this high-quality spectrogram obtained at the period of enhanced accretion is still relevant and has not been surpassed for revealing physical properties of accretion/outflow activity in RZ Psc system. 

The spectrogram was obtained on the night of November 16, 2013 with the Keck I telescope and HIRES echelle spectrograph (\textit{PI}: B.Zuckerman). Observations were carried out with 1.148 arcsec projected slit width resulted in nominal spectral resolution $R\approx38000$. The wavelength coverage of the spectrograms was $\Delta \lambda\approx4700-9000$~\AA\ with the signal-to-noise ratio of about $S/N\approx120$ (per pixel) in the region near $\mathrm{H\alpha}$ line. Details on the data processing are given in \citet{potravnovFlaresAccretionActivity2019}. The processing workflow including standard calibration routine for the science frames, 1D spectrum extraction and wavelength calibration were made with the \textsc{Makee}\footnote{https://sites.astro.caltech.edu/~tb/makee/} software (written by T.Barlow). The heliocentric corrections was applied to the wavelength scale and afterwards it was shifted for the the stellar radial velocity \(RV=-1.3\pm0.3\) km s\(^{-1}\) \citep{Potravnov_2014}. Thus, hereafter we used the spectrum in the rest frame associated with the star. The flux normalization was performed using the approximation of the continuum level with the low-order cubic spline. 

\begin{figure}
    \centering
    \includegraphics[width=\columnwidth]{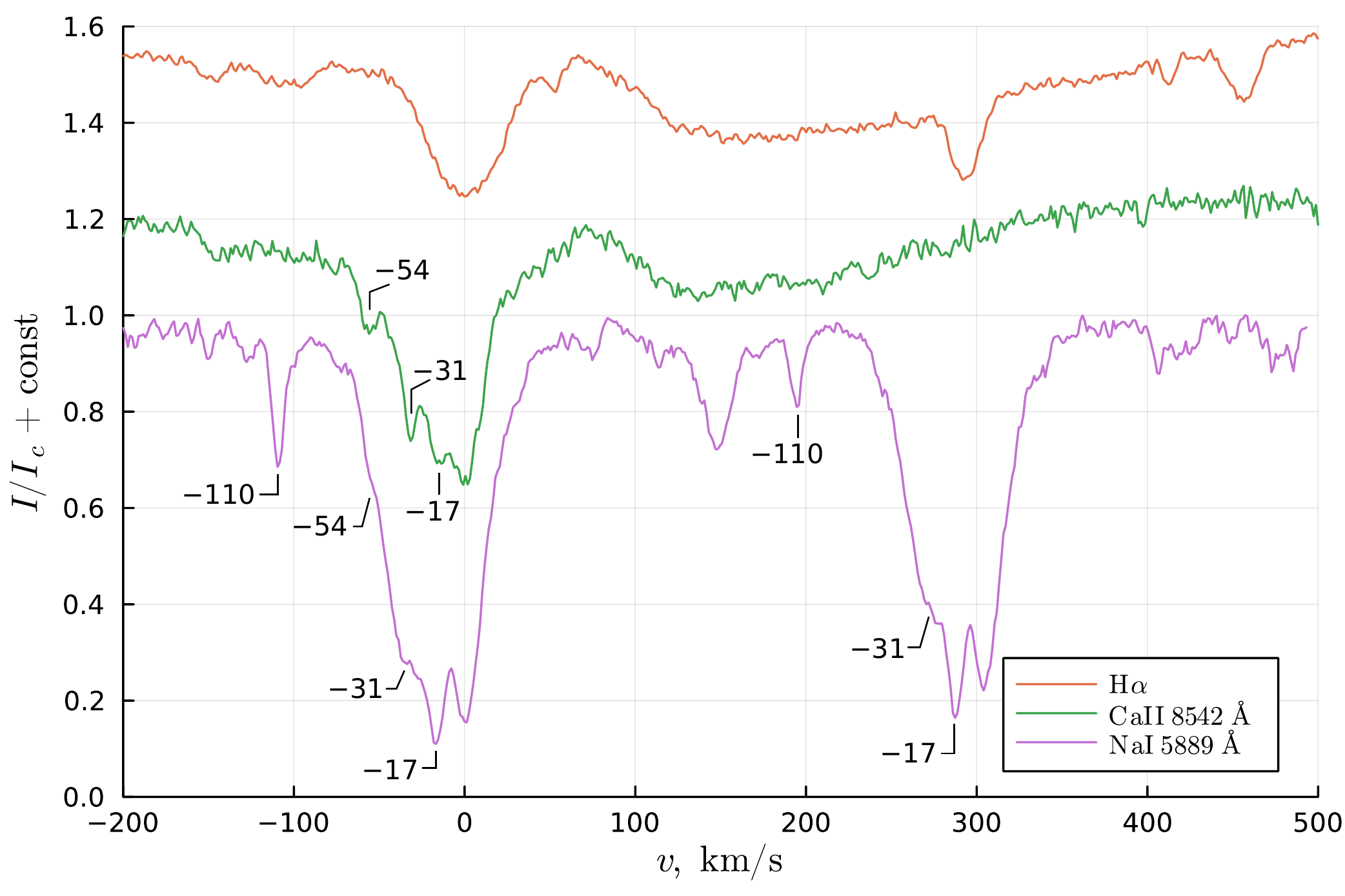}
    \caption{$\mathrm{H\alpha}$, CaII and NaI lines observed at November 16, 2013 in RZ Psc spectrum. The BACs, noticeable in \ion{Ca}{II} 8542 \AA\ and \ion{Na}{I} 5889\ \AA\ profiles, are labeled on the plot.}
    \label{fig:CaNa-observed}
\end{figure}

The Nov. 16 spectrogram was obtained soon after the deep minimum when the star was close to its normal brightness. The $\mathrm{H\alpha}$ and \ion{Ca}{II} lines demonstrated clear inverse P Cyg profile, with filled-in photospheric component and broad redshifted absorption extended up to +580 km s\(^{-1}\). BACs in sodium lines are clearly seen in Fig. \ref{fig:CaNa-observed} and also traceable in profiles of \ion{Ca}{II} lines.  No corresponding wind features were observed in the $\mathrm{H\alpha}$ line  (Fig.~\ref{fig:CaNa-observed}). 

\section{Model}\label{sec:model}

In our modeling we use the following parameters of RZ Psc: \(R_\star = 1.2\ \mathrm{R_\odot}\), \(M_\star = 1.1\ \mathrm{M_\odot}\) \citep{potravnovFlaresAccretionActivity2019} and \(T_\star = 5350\) K \citep{Potravnov_2014}. Here only a brief description of the model is given, where the focus was shifted on the details most relevant for this work. The more detailed description can be found in our previous works \cite{dmitrievFormationHydrogenEmission2019} and \cite{dmitrievModelsMagnetosphericAccretion2022}. 

\begin{figure}
    \centering
    \includegraphics[width=\columnwidth]{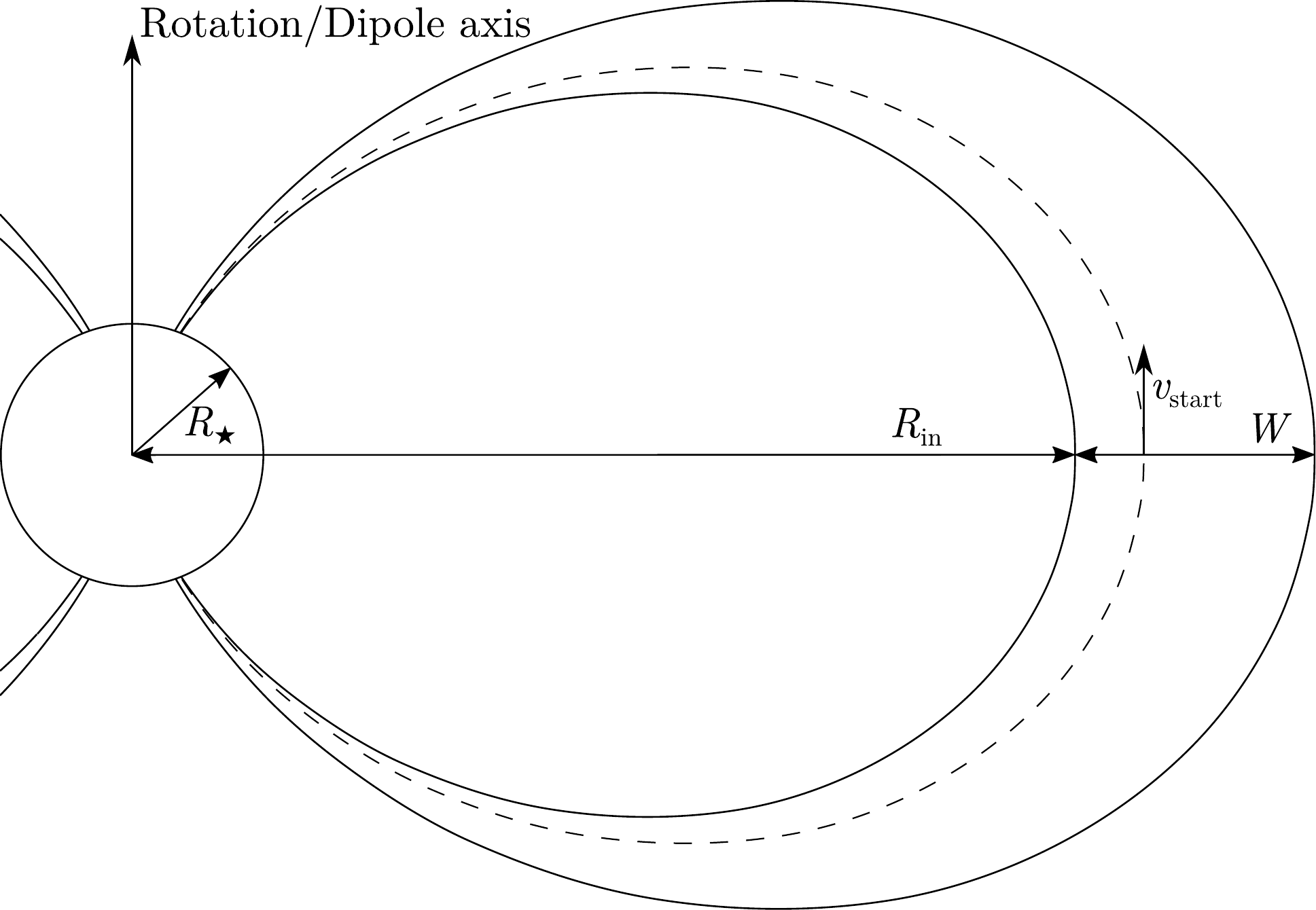}
    \caption{The schematic of the magnetosphere.}
    \label{fig:mag-scheme}
\end{figure}

In the framework of the classical approach, the magnetosphere is assumed to be formed by the dipole field aligned with the stellar rotation where the gas falls freely along magnetic field lines. In the scope of these assumptions it has 5 independent parameters: accretion rate $\dot{M}$, maximum temperature in the magnetosphere $T_\mathrm{max}$, inner radius $R_\mathrm{in}$, width $W = R_\mathrm{out} - R_\mathrm{in}$, where \(R_\mathrm{out}\) is the outer radius of the magnetosphere and the initial velocity \(v_\mathrm{start}\) which should be close to the thermal velocity \(v_\mathrm{th}\). Following \cite{hartmannMagnetosphericAccretionModels1994} and \cite{muzerolleEmissionLineDiagnosticsTauri2001} we set \(v_\mathrm{start}\) to 10 km/s. From these parameters density, temperature and motion of the gas are completely determined (see, for example, \cite{hartmannMagnetosphericAccretionModels1994}). In  Fig. \ref{fig:mag-scheme} the schematic of the magnetosphere is given.

We assume purely hydrogen gas and write state equations in the form
\begin{equation}\label{eq:conservation}
    \nabla\cdot(\mathbf{v}n_i) = \sigma_i\ \mathrm{for}\ i \ge 1,
\end{equation}
with constraint equation
\begin{equation}\label{eq:constraint}
n_\mathrm{H} = n_e + \sum\limits_{i=1}^\infty n_i,    
\end{equation}
where \(n_i\) are level populations, \(n_e\) is electron concentration and \(n_\mathrm{H}\) is total hydrogen concentration and $\sigma_i$ are sources and sinks for $i$-th hydrogen level:
\begin{equation}\label{eq:sources}
\begin{split}
\sigma_i =  &\sum\limits_{k=i+1}^\infty n_k (A_{ki} + B_{ki}J_{ki}) + \sum\limits_{j=1}^{i-1} n_jB_{ji}J_{ij} + n_e \sum\limits_{j\neq i}^\infty n_jq_{ji} -  \\ 
&n_i \left[ \sum\limits_{j=1}^{i-1} (A_{ij} + B_{ij}J_{ij}) + \sum\limits_{k=i+1}^\infty B_{ik}J_{ik} + n_e\sum\limits_{j \neq i}^\infty q_{ij}\right] + \\ 
&n_e^2(C_i + B_{ci})  + n_e^3Q_{ci} - n_iB_{ic} - n_i n_e q_{ic}, \ \ i=1,\ 2,\ ... \\
\end{split}
\end{equation}
The terms of Eq. \eqref{eq:sources} include radiative transitions (Einstein's coefficients \(A_{ij}\), \(B_{ij}\)), transitions induced by collisions with free electrons ($q_{ij}$), spontaneous and induced by radiation or electron collisions recombinations ($C_i$, $B_{ci}$ and $Q_{ci}$) and radiative and collisional ionizations ($B_{ic}$ and $q_{ic}$). 
The sums are truncated at 15-th level. The mean intensities \(J_{ij}\) are computed using the Sobolev's approximation \citep{sobolevMovingEnvelopesStars1960, grachevAnalysisLineProfiles1975, rybickiGeneralizationSobolevMethod1978}. 

The radiative ionization and induced recombination coefficients (\(B_{ic}\) and \(B_{ci}\)) are computed as follows
\begin{equation}
    B_{ic} = 4\pi\int\limits_{\nu_{ci}}^\infty \alpha_{ic}(\nu)\frac{J_\nu}{h\nu} d\nu
\end{equation}
\begin{equation}
    B_{ci} = \frac{i^2h^3}{(2\pi m_ek_BT)^{3/2}}e^{\frac{h\nu_{ci}}{k_BT}}4\pi\int\limits_{\nu_{ci}}^\infty \alpha_{ic}(\nu) \frac{J_\nu}{h\nu} e^{-\frac{h\nu}{k_BT}} d\nu,
\end{equation}
where \(\alpha_{ic}\) is photoionization cross section from level \(i\), \(\nu_{ic}\) is the threshold frequency for level \(i\), \(T\) is the local temperature in the magnetosphere and \(J_\nu\) is the mean intensity in the continuum. We neglect the continuum radiation from the magnetosphere, and assume that the external sources of radiation are the blackbody radiation of the star and the accretion spot at the base of the magnetosphere 
\begin{equation}
    J_\nu = W_\star B_\nu(T_\star) + W_\mathrm{spot}B_\nu(T_\mathrm{spot}),
\end{equation}
where \(B_\nu\) is the Planck's law, \(W_\star\) and \(W_\mathrm{spot}\) are the geometrical dilution factors for the star and for the accretion spot and \(T_\mathrm{spot}\) is the temperature of the accretion spot. This temperature is computed from the assumption that all of the kinetic energy of the falling gas at the base of the magnetosphere is radiated away as blackbody radiation \citep{hartmannMagnetosphericAccretionModels1994}.

In present paper we take into account the advective term $\nabla\cdot(\mathbf{v}n_i)$ in equations \eqref{eq:conservation}, because it can be significant for low accretion rates, observed in RZ Psc. Its importance in T Tau stars magnetospheres was first stated by \cite{martinThermalStructureMagnetic1996}, and the impact on the emission spectrum was first considered by \cite{dmitrievModelsMagnetosphericAccretion2022}.

To simplify the system of partial differential equations \eqref{eq:conservation} we neglect advective term $\nabla\cdot(\mathbf{v}n_i)$ for excited levels \(i > 1\). This assumption is based on observation that populations of excited levels are largely controlled by spontaneous deexcitation and the time scale of this process is much smaller than kinematic timescale of $\nabla\cdot(\mathbf{v}n_i)$. The first level is mostly controlled by spontaneous recombinations with much larger timescale. We also assume that populations of excited levels are much smaller than electron concentration $n_e$ and first level population $n_1$. Thus, the system of equation and the constraint equation \eqref{eq:constraint} becomes
\begin{equation}\label{eq:main-popul}
\begin{cases}
\sigma_1& = \nabla\cdot(\mathbf{v}n_1)\\ 
\sigma_i& = 0\ \mathrm{for}\ i > 1 \\
n_e& = n_\mathrm{H} - n_1.
\end{cases}
\end{equation}
The initial conditions are obtained by solving the equations
\[
\sigma_i = 0,\ i \ge 1
\]
at the beginning of each stream line.

The line profile is computed using ray-by-ray integration of radiation transfer equation \citep{muzerolleEmissionLineDiagnosticsTauri2001}. The absorption coefficient is computed using Doppler profile, and full frequency redistribution is assumed. Here another important parameter arises: angle $i$ between line of sight and the axis of the magnetosphere. 

Because the observed profile is weak (see Fig. \ref{fig:CaNa-observed}), it is impossible to separate magnetosphere component from the spectrum of the photosphere with sufficient precision. So, for each ray that passes through the surface of the star we put the photosphere spectrum \(I^\star_\nu\), shifted to correct for the solid-body rotation of the star, in the equation for ray intensity
\begin{equation}
    I_\nu = I^\star_\nu e^{-\tau_\nu} + \int\limits_0^{\tau_\nu} S_\nu e^{\tau} d\tau.
\end{equation}
Here $\tau_\nu$ is optical depth along the line of sight and $S_\nu$ is the source function.

Subsequently, five parameters are needed to compute the line profile, excluding the parameters of the star:  accretion rate $\dot{M}$, maximum temperature in the magnetosphere $T_\mathrm{max}$, inner radius of the magnetosphere $R_\mathrm{in}$, width of the magnetosphere $W$ and angle $i$. 

\section{Fitting procedure and results}\label{sec:res+fit}

\begin{table}
	\centering
	\caption{Parameter grid.}
	\label{tab:parameter}
	\begin{tabular}{lcccl} 
		\hline
		Parameter & Minimum value & Step & Maximum value & Units\\
		\hline
		$\log\dot{M}$ & -8.4 & 0.2 & -11 & $\mathrm{M}_\odot/\mathrm{yr}$\\
		$T_\mathrm{max}$  & 7000 & 1000 & 15000 & K\\
		$R_\mathrm{in}$& 2 & 1 & 10 & $R_\star$\\
		$W$ & 1 & 0.2 & 4 & $R_\star$ \\
		$i$ & 35 & 5 & 60 & Degrees  \\
		\hline
	\end{tabular}
\end{table}

We used the grid of the parameters with a total number of 108864 points described in Table \ref{tab:parameter}. Because the magnetospheric accretion is impossible at the distances larger then the corotation radius $R_\mathrm{cor}$, the models where outer radius $R_\mathrm{out} = R_\mathrm{in} + W$ exceeded $11 R_\star$ were disregarded, leaving a total number of 86184 computed profiles. This estimate of the corotation radius was calculated assuming that the rotational velocity on the equator \(v_\mathrm{eq}\) is equal to \(v\sin i = 12\ \mathrm{km\ s^{-1}}\) obtained from the spectrum, because the inclination angle $i$ is unknown. The true value of corotation radius \(R_\mathrm{cor}\) must be \(\le 11\ \mathrm{R_\star}\). 

For each obtained profile the residual with observations $\delta$ was computed
\begin{equation}\label{eq:residual}
    \delta = \sqrt{\frac{1}{N_\mathrm{freq}}\sum\limits_{|v| \ge 30\ \mathrm{km\ s^{-1}}}^{N_\mathrm{freq}} (r_\mathrm{mod}(\nu) - r_\mathrm{obs}(\nu))^2},
\end{equation}
where \(r_\mathrm{mod} = I_\mathrm{mod}/I_c\) is the computed profile, \(r_\mathrm{obs} = I_\mathrm{obs}/I_c\) is the observed profile and \(N_\mathrm{freq}\) is the number of frequencies where the profile was computed.
The central part of the profile where \(|v| = c|\nu - \nu_{\mathrm{H\alpha}}|/\nu_{\mathrm{H\alpha}} > 30\ \mathrm{km\ s^{-1}}\) is removed from the sum because our model cannot produce strong enough emission at this region. Similar to \cite{thanathibodeeVariableAccretionProtoplanet2020}, we add a central Gaussian component
\begin{equation}\label{eq:central-comp}
    r_\nu = r_\mathrm{mod} + A\exp\left(-\frac{c^2(\nu - \nu_{\mathrm{H\alpha}})^2}{2\nu_{\mathrm{H\alpha}}^2\Delta v^2}\right),
\end{equation}
to remove this difference. Its parameters \(A,\ \Delta v\) are computed by fitting $r_\mathrm{mod}(\nu) - r_\mathrm{obs}(\nu)$. This central component may originate in an accretion shock at the base of the magnetosphere \citep{dodinNonLTEModelingStructure2015, dodinStructureSpectrumAccretion2018} or in active regions in the chromosphere of the star. The existence of these regions is supported by the X-ray activity of RZ Psc \citep{punziYoungStarRZ2018}. We emphasize again, that this narrow central component was omitted in the computation of residuals.

\begin{figure}
	\includegraphics[width=\columnwidth]{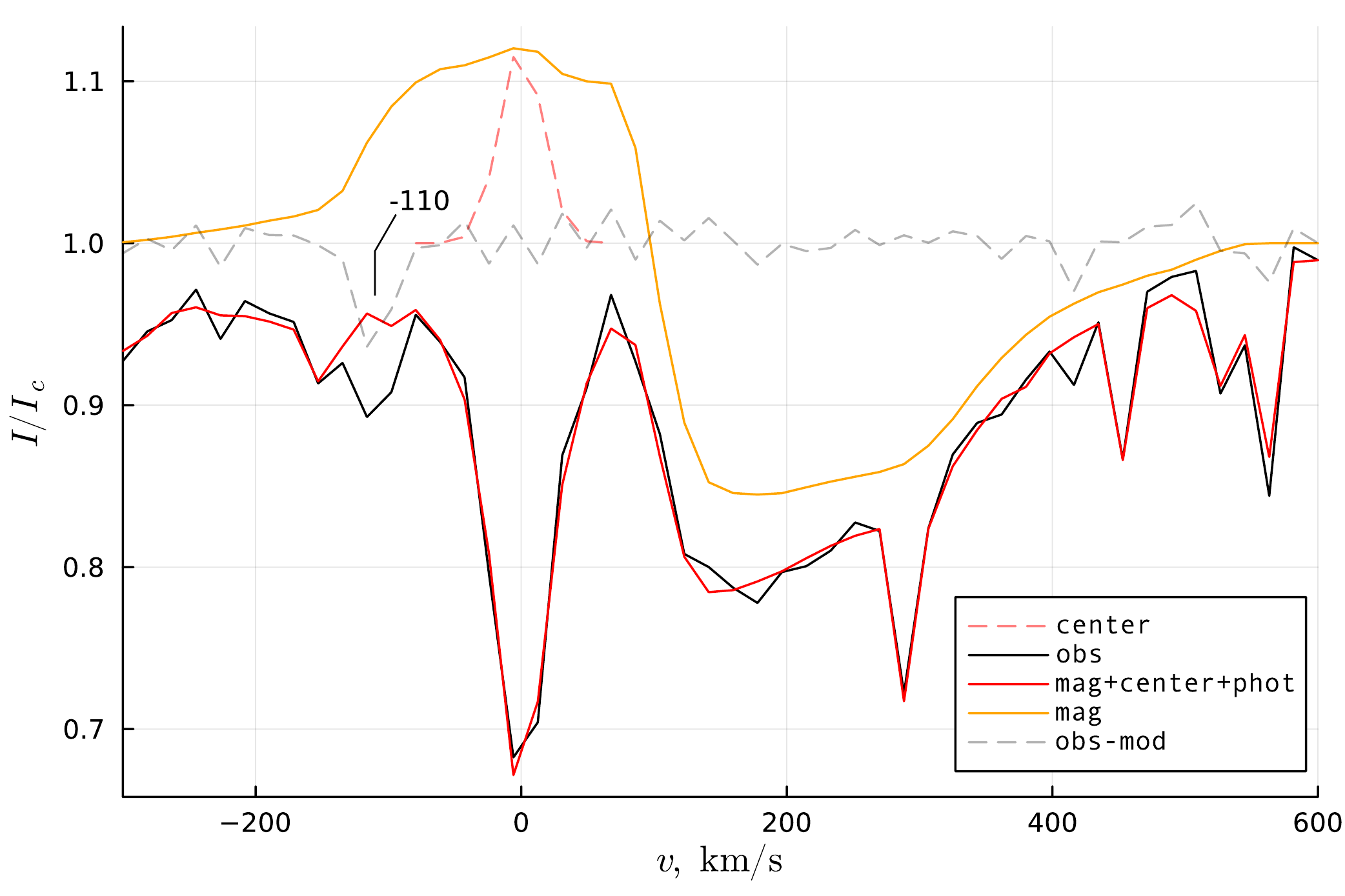}
    \caption{One of the computed \(\mathrm{H\alpha}\) profiles (red line labeled \texttt{mag+center+phot}) with $\delta < \sqrt{2}\delta_\mathrm{min}$ in comparison with observations (black line labeled \texttt{obs}). Magnetosphere only profile (without photosphere, yellow line labeled \texttt{mag}), central component (red dashed line labeled \texttt{center}) and deviation from observation (gray dashed line labeled \texttt{mod-obs}) are also shown. The model parameters are: \(\dot{M} = 2.5\times10^{-10}\ \mathrm{M_\odot/yr}\), \(T_\mathrm{max} = 9000\) K, \(R_\mathrm{in} = 5\ \mathrm{R_\star}\), \(W = 2\ \mathrm{R_\star}\), \(i = 45^\circ\). The central component has \(A = 0.12\) and \(\Delta v = 16\) km/s. The significant deviation from the observed profile at \(v \approx -110 \mathrm{km\ s^{-1}}\) is marked.}
    \label{fig:profile-ex}
\end{figure}

Fig. \ref{fig:profile-ex} shows an example of theoretical profile constructed from the different components in comparison with observed one. We found, that magnetosphere accretion model with weak central peak can explain virtually the entirety of the observed profile. The only discrepancy worth of discussion is positioned at \(v \approx -110\ \mathrm{km\ s^{-1}}\) as can be seen in Fig. \ref{fig:profile-ex}. This velocity coincides with one of the BACs observed in NaI lines (see Fig. \ref{fig:CaNa-observed}), so this may be an \(\mathrm{H\alpha}\) absorption from the same outflowing gas.

We derive observed profile parameters by computing mean of parameters of the models with sufficiently low residuals $\delta < \sqrt{2}\delta_\mathrm{min}$, where \(\delta_\mathrm{min}\) is minimum value of \(\delta\) on the grid. For error estimate we use standard deviation \(\sqrt{D}\) where \(D\) is the dispersion of such parameters. However, its important to highlight the limitations of this approach. If, for example, for one of the parameters the inequality $\delta < \sqrt{2}\delta_\mathrm{min}$ is true for all grid points, this approach will yield the midpoint as the result and the square root of the dispersion of the grid points \(\sqrt{D_\mathrm{grid}}\) as an error. Both those values are completely independent of observations, so, in reality, this parameter is undetermined. To highlight such situations we compute the confidence relation $\sqrt{D_\mathrm{grid}}/\sqrt{D}$ for each of the parameters. If this relation is close to unity the parameter is considered unconstrained. In our case such situation occurs for two parameters of the magnetosphere: width $W$ and maximum temperature $T_\mathrm{max}$. We chose ranges for these parameters that are in agreement with the results of theoretical works (see \cite{hartmannMagnetosphericAccretionModels1994}, \cite{muzerolleEmissionLineDiagnosticsTauri2001}, \cite{Lima2010}). 

We use one of NaI optical doublet component (5890\ \AA) and one of CaII infrared triplet component (8542\ \AA) to restrict the model parameters. In the RZ Psc spectrum observed at November 16 2013 there is no noticeable absorption in the red wing of NaI 5890 \AA, but there is a profound absorption feature in CaII 8542\AA\ from \(\sim\)50 to \(\sim\)500 \(\mathrm{km\ s^{-1}}\) very similar to absorption feature observed in \(\mathrm{H\alpha}\) (see Fig. \ref{fig:CaNa-observed}). We argue that this is due to the fact that the magnetosphere is optically thin in NaI 5890 \AA, but optically thick in CaII 8542\AA, and we can use this fact to disregard some of the models with small \(\delta\). To achieve this we computed absorption coefficients in these lines for the conditions arising in the magnetosphere using Cloudy \citep{ferland2017ReleaseCloudy2017a} package and calculated absorption profiles of the magnetosphere with \(\delta < \sqrt{2}\delta_\mathrm{min}\) at 200 km/s. Then we rejected the models for which this value was smaller than \(0.9I_c\) for NaI 5890\ \AA\ or bigger than \(0.9I_c\) for CaII 8542\ \AA. 

\begin{figure}
    \centering
    \includegraphics[width=\columnwidth]{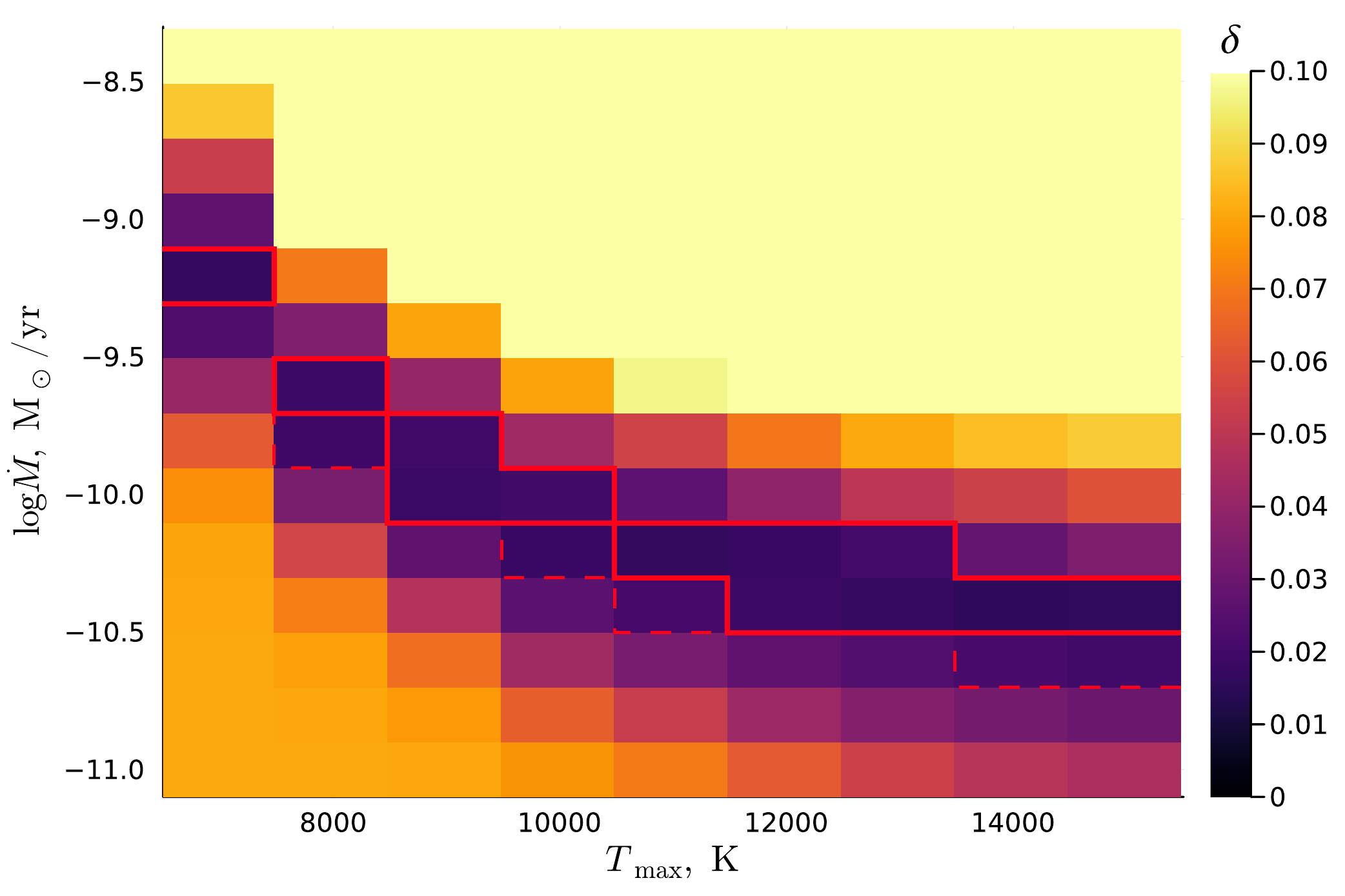}
    \caption{Minimum residual value \(\delta\) for computed models in the stationary case with fixed \(\dot{M}\) and \(T_\mathrm{max}\). The residual value is shown in color. The region where \(\delta < \sqrt{2}\delta_\mathrm{min}\) and absorption in CaII and NaI lines satisfies our criteria lies inside the red border. The dashed red border separates models where \(\delta < \sqrt{2}\delta_\mathrm{min}\) but the criteria is not satisfied.}
    \label{fig:statMT}
\end{figure}

\begin{figure}
    \centering
    \includegraphics[width=\columnwidth]{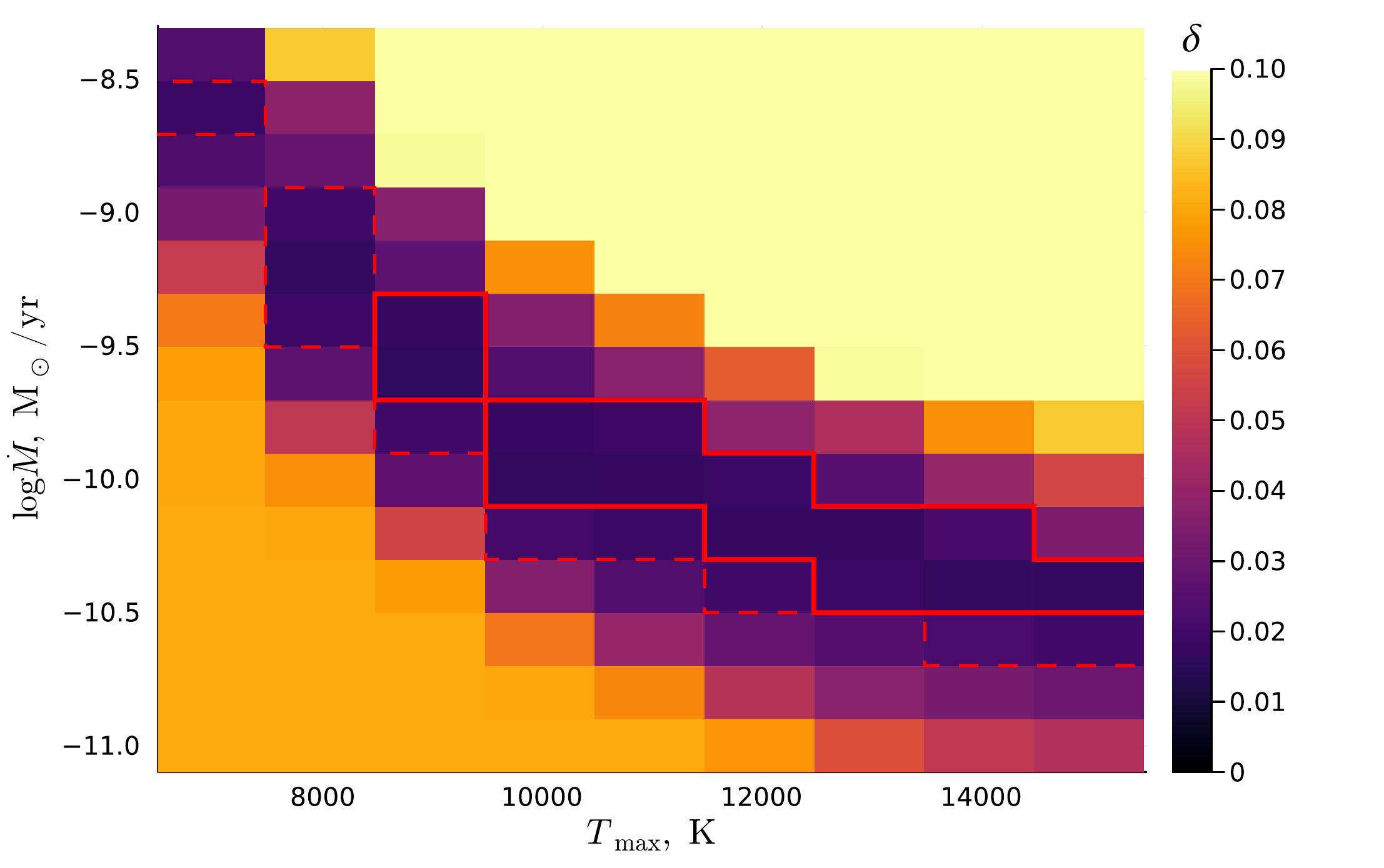}
    \caption{Minimum residual value \(\delta\) for computed models with fixed \(\dot{M}\) and \(T_\mathrm{max}\) with advective term \(\nabla\cdot(vn_1)\) taken into account. The residual value is shown in color. The region where \(\delta < \sqrt{2}\delta_\mathrm{min}\) and absorption in CaII and NaI lines meets our criteria lies inside the red border. The dashed red border separates models where \(\delta < \sqrt{2}\delta_\mathrm{min}\) but the criteria is not satisfied.}
\label{fig:nonstatMT}
\end{figure}


To illustrate the effect of advective term \(\nabla\cdot(\mathbf{v}n_1)\) in equations \eqref{eq:main-popul} we also calculated profiles in the stationary assumption
(\(\sigma_i = 0\), see section \ref{sec:model}). Fig. \ref{fig:statMT} shows how the residual \(\delta\) depends on accretion rate \(\dot{M}\) and temperature
\(T_\mathrm{max}\) in the stationary assumption. Fig. \ref{fig:nonstatMT} is the same, but with the advective term taken into account. The minimum residual value \(\delta_\mathrm{min}\) is approximately 0.017 for both cases. It can be seen clearly
that stationary assumption produces smaller accretion rates for low temperatures, although the difference is only about half an order. This agrees with theoretical results described in \cite{dmitrievModelsMagnetosphericAccretion2022}.
This difference allows us to disregard temperatures smaller than 9000~K using CaII and NaI lines (in stationary assumptions these temperatures are valid, see Fig. \ref{fig:statMT}). 
Subsequently, this has a significant impact on the precision of obtained accretion rate, as for low temperatures accretion rates up to \(10^{-9}\ \mathrm{M_\odot yr^{-1}}\) are required to produce observed \(\mathrm{H\alpha}\) profile.

The obtained average values of parameters are presented in Table~\ref{tab:fin-res}. The accretion rate \(\dot{M}\), inner radius of the magnetosphere \(R_\mathrm{in}\) and inclination angle \(i\) are well determined with confidence relation \(\sqrt{D_\mathrm{grid}}/\sqrt{D} \approx 3\). However, for temperature \(T_\mathrm{max}\) and magnetosphere width \(W\) this relation is close to 1. This is due to the fact that those two parameters are only bounded from bellow on the grid, as can be seen in Fig. \ref{fig:nonstatMT}. However, the width \(W\)  cannot be much larger, because the outer radius \(R_\mathrm{out} = R_\mathrm{in} + W\) cannot exceed the corotation radius.

\begin{table}
	\centering
	\caption{Results with advection taken into account. Parameters in red rows are unconstrained due to their low confidence relation.}
	\label{tab:fin-res}
	\begin{tabular}{lclcl}
		\hline
		Parameter & Value & Error & Confidence & Units\\
		\hline
		$\log\dot{M}$ & -10.1 & $\pm0.3$ & 3.0 & $\mathrm{M}_\odot/\mathrm{yr}$\\
		\rowcolor{RedAlert}
		$T_\mathrm{max}$ & $12500$ & $\pm2100$ & 1.4 & K\\
		$R_\mathrm{in}$& 5.5 & $\pm0.9$ & 3.0 & R$_*$\\
		\rowcolor{RedAlert}
		$W$ & 3.0 & $\pm0.6$ & 1.9 & R$_*$ \\
		$i$ & 43 & $\pm3$ & 3.8 & Degrees  \\
		\hline
	\end{tabular}
\end{table}





\section{Discussion}\label{sec:disscuss}
According to results of our modeling the logarithm of mass accretion rate onto RZ Psc during the "flare" of its accretion activity on 2013 Nov. 16 was \(\log\dot{M} = -10.1\pm0.3\) (\(\dot{M} \approx 7\times10^{-11} \mathrm{M}_\odot\mathrm{yr^{-1}}\)) that is, about 10 times more than before the burst.
This suggests that the accretion process at the late stages of the Pre-Main Sequence evolution is extremely unsteady. But even at the moment of the maximal accretion activity the accretion rate onto RZ Psc was very small compared to typical rates of T Tauri stars \(10^{-8} - 10^{-7}\ \mathrm{M_\odot\ yr^{-1}}\). This is probably one of the reasons for sustained accretion activity in $\approx$20 Myr old RZ Psc system. The other reason for existence of the long living disk around RZ Psc is the operation of accretion process in the weak magnetic propeller mode \citep{grininMagneticPropellerEffect2015}. It explains the very interesting property of this star outside of rare accretion bursts: existence of the spectroscopic signatures of the matter outflow and lack of any signs of accretion. In the paper cited above we argued that the terminal velocity of the expelled gas does not exceed the local escape velocity. \cite{romanova2018PropertiesStrongWeakPropeller} called such a mode of accretion as the "soft" propeller. In this case the magnetosphere works as a mixer. It is a very economical mode of accretion when the CS gas is expelled from the star and return back into the disk. Such a disk can survive during a very long time.

The weak accretion rate in the RZ Psc system indicates, that the ionization in the falling gas can deviate from equilibrium. In the case of RZ Psc accounting of this effect allows us to reject models with low temperature using CaII and NaI lines and determine the accretion rate and other parameters more precisely. This result demonstrates importance of the temperature diagnostic for the modeling of magnetospheres in young stars).

Using the obtained values of \(\dot{M}\) and \(R_\mathrm{in}\) one can estimate the strength of the dipole component of the magnetic field on the equator of RZ Psc \(B_\mathrm{dip}\). Assuming that \(R_\mathrm{in}\) is the truncation radius we can rewrite equation (2.2) from \cite{bouvierMagnetosphericAccretionClassical2007} as
\begin{equation}\label{eq:trunc}
    B_\mathrm{dip} = \left(\frac{R_\mathrm{in}}{7.1\ \mathrm{R_\star}}\right)^{7/4}\left(\frac{\dot{M}}{10^{-8}\ \mathrm{M_\odot/yr}}\right)^{1/2}\left(\frac{M_\star}{0.5\ \mathrm{M_\odot}}\right)^{1/4}\left(\frac{R_\star}{2\ \mathrm{R_\odot}}\right)^{-5/4}\ \mathrm{kGs}.
\end{equation}
Substituting values of \(\dot{M}\) and \(R_\mathrm{in}\) from Table \ref{tab:fin-res} we obtain 
\[
B_\mathrm{dip} = \left(\frac{5.5}{7.1}\right)^{7/4}\left(\frac{10^{-10.1}}{10^{-8}}\right)^{1/2}\left(\frac{1.1}{0.5}\right)^{1/4}\left(\frac{1.2}{2}\right)^{-5/4} \approx 0.13 \pm 0.08\ \mathrm{kGs}.
\]
This value is significantly lower than the typical value (\(B \approx 1\) kGs) observed for T Tauri stars. 

From the Table 2 we have outer radius of the magnetosphere $R_\mathrm{out} = R_\mathrm{in} + W \approx 8.5 \pm 1.5\ R_\star$. According to \citep{grininMagneticPropellerEffect2015} the corotation radius of RZ Psc is about 8-9 $R_\star$. This value coincides with our estimation of the outer radius of the magnetosphere $R_\mathrm{out}$. Therefore our estimate of the magnetic field admits existence of matter outflow in the magnetic propeller mode from the outer regions of the magnetosphere. This explains the presence of both accretion and outflow signatures in the spectrum (see Fig. \ref{fig:CaNa-observed}). But, if we put the accretion rate observed outside the accretion burst \(\dot{M} = 7\times10^{-12}\ \mathrm{M}_\odot\mathrm{yr^{-1}}\) and the estimated magnetic field (\(\approx 0.1\) kGs) in the equation \eqref{eq:trunc} then the inner radius of the magnetosphere will extend to $\approx 10\ R_\star$. This value is larger then the corotation radius, which explains the absence of accretion signatures and existence of only outflow signatures in the spectra obtained in the normal state of the star.

In our calculations we used the classical model of the stellar magnetosphere based on the dipole magnetic field.  The recent observations of magnetic fields in the WTTS's demonstrate the large diversity in strengths and topology of the large-scale magnetic field \citep{Donati2011MagneticFieldOfTWHya, Donati2014ModelingMageticActivityWTTSLkCa4, Donati2017HotJupiterMagneticActiveStarV830Tau, Hill2017MagneticActivityWTTSPar1379Par2244, Hill2019MagneticTopologyWTTSTWA6TWA8A, Nicholson2018magneticActivityWTTSTWA9AV1095Sco, Yu2017HotJupiterActiveTTauTAP26}. In the light of this the direct measurements of magnetic field in RZ Psc are highly desirable.

From the point of view of the variable CS extinction model, the inclination angle $i$ is one of the key parameters of CS disks. Our modeling showed that the inclination angle of RZ Psc  $i = 43\,\pm\,3^\circ$. This value is smaller in comparison with the inclination angle \(i\approx70^\circ\) of the photometrically active UXOrs  \citep{kreplinRevealingInclinedCircumstellar2013, kreplinResolvingInnerDisk2016, pontoppidan2007ModelingSpitzerObservationsVVSerII, Langlois2018ScatteredLightTransitionDiskRYLupi}, and this difference is probably the main reason of the low photometric variability of RZ Psc. 

\section{Conclusions}\label{sec:concl}

In this paper we modeled \(\mathrm{H\alpha}\) emission in the spectrum of RZ Psc during the accretion burst in November 2013 using a magnetosphere model described in \cite{dmitrievFormationHydrogenEmission2019} and \cite{dmitrievModelsMagnetosphericAccretion2022}. The main results can be summarized as follows:
\begin{enumerate}
    \item The accretion rate increased approximately by an order of magitude to the value of \(\log \dot{M} = -10.1\pm0.3\), that corresponds to \(\dot{M} = 7\times10^{-11} \mathrm{M}_\odot\mathrm{yr^{-1}}\). Outside the episode of the accretion burst, the accretion rate is too small to produce any noticable accretion signatures.  
    \item The inclination angle \(i = 43\,\pm\,3^\circ\) is low compared to the typical one for UX Ori stars \(i \approx 70^\circ\), which can be a reason of the low photometric variability of the star.
    \item The accounting for advective effects allowed us to place a lower limit on the temperature in the magnetosphere at $\approx 10000$ K using observed profiles of the IR CaII triplet lines and D Na I resonance lines, which significantly improved precision of our estimate of accretion rate.
    \item The magnetosphere extends approximately to the corotation radius. Thus, at the outermost regions the magnetic field can expel some of the accreting gas. This explains the presence of BACs attributed to the magnetic propeller in the November 16 spectrum observed during the accretion burst.
    \item We estimate the dipole magnetic field component as \(B_\mathrm{dip} \approx 0.1\) kGs using obtained values of accretion rate and inner radius of the magnetosphere. This value is quite low for T Tauri stars. In this regard it would be interesting to directly measure the magnetic field of RZ Psc.
\end{enumerate}
 
\section*{Acknowledgements}
The authors thank the referee for useful suggestions that helped to improve the manuscript.
DVD, TAE and VPG acknowledge the support of Ministry of Science and Higher Education of the Russian Federation under the grant no. 075-15-2020-780 (N13.1902.21.0039).
This research has made use of the Keck Observatory Archive (KOA), which is operated by the W. M. Keck Observatory and the NASA Exoplanet Science Institute (NExScI), under contract with the National Aeronautics and Space Administration.\\

\section*{Data Availability}

All data used in this article will be shared on reasonable request to the corresponding author.
 



\bibliographystyle{mnras}
\bibliography{example} 








\bsp	
\label{lastpage}
\end{document}